\documentclass[pre,showpac,preprintnumbers,twocolumn]{revtex4}
\usepackage{epsfig}
\usepackage{amssymb,amsmath,amsthm}

\begin{document}
\title{Localized states in bistable pattern forming systems}
\author{U. Bortolozzo$^{1,2}$, M. G. Clerc$^{3}$, C. Falcon$^{3}$, S.
Residori$^{1}$, and R. Rojas$^{1}$} \affiliation{
$^{1}$Institut Non Lin\'{e}aire de
Nice, UMR 6618 CNRS-UNSA, 1361Route des
Lucioles, F-06560 Valbonne, France. \\
$^{2}$Istituto Nazionale di Ottica Applicata, Largo E. Fermi 6 50125 Florence, Italy
\\
$^{3}$Departamento de F\'{i}sica, Facultad de Ciencias
F\'{i}sicas y Matem\'{a}ticas, Universidad de Chile, Casilla
487-3, Santiago, Chile.}

\begin{abstract}
We present an unifying description close to a spatial
bifurcation of localized states, appearing as
large amplitude peaks nucleating over a pattern of lower
amplitude. Localized states are pinned over a lattice
spontaneously generated by the system itself. We show that the
phenomenon is generic and requires only the coexistence of two
spatially periodic states. At the onset of the spatial
bifurcation, a forced amplitude equation is derived for the
critical modes, which accounts for the appearance of localized
peaks.
\end{abstract}


\maketitle

During the last years emerging localized structures in dissipative
systems have been observed in different fields, such as domains in
magnetic materials \cite{Eschenfelder}, chiral bubbles in liquid
crystals \cite {Oswald}, current filaments in gas discharge
experiments \cite{Astrov97}, spots in chemical reactions
\cite{Swinney}, localized 2D states  in fluid surface waves
\cite{Edwards}, oscillons in granular media \cite{Melo}, isolated
states in thermal convection \cite{Heinrichs}, solitary waves in
nonlinear optics \cite{Newell}, just to mention a few. In
one-dimensional systems, localized patterns can be described as
homoclinic  orbits passing close to a spatially oscillatory state
and converging to an homogeneous state \cite{Cross,Coullet2002},
whereas domains are seen as heteroclinic trajectories joining the
fixed points of the corresponding dynamical system
\cite{VanSaarlos}. Recently, in a nematic liquid crystal light valve
with optical feedback it has been found experimentally a different
type of localized states, appearing as a large amplitude peaks
nucleating over a lower amplitude pattern and therefore called {\it
localized peaks} \cite{Bortolozzo2005}. Similar observations have
been reported in a Newtonian fluid when non linear surface waves are
parametrically excited with two frequencies \cite{Fineberg} and in
numerical simulations of an atomic vapor with optical feedback
\cite{atomic}.  Recently, longitudinal modes with localized peaks
over a spatially modulated background have been shown in numerical
simulations of Maxwell-Bloch equations for a semiconductor laser
\cite{Lionel}.

 All these different types of localized states appear over a
patterned background and thus constitute a different class of
structures with respect to the ones appearing over an uniform
background. The aim of this manuscript is to show that localized
peaks are a generic class of localized states, appearing whenever a
pattern forming system exhibits coexistence of two spatially
periodic states. The mechanisms that originate this circumstances
are more than a few, for instance, one can consider a multi-stable
system, which shows two consecutive spatial bifurcations to
different states when one parameter is changed.  There is a large
number of physical systems that display this kind of behavior,
therefore there is a vast number of possible models. In order to
derive an unifying and simple description of localized peaks, we
develop a theoretical model for one-dimensional spatially extended
systems close to a spatial bifurcation. The model, which shows
coexistence between different patterns and stable front solutions
between them, is based on an amplitude equation that includes a
spatial parametric forcing. This extension with respect to
conventional amplitude equations, allows to describe localized
patterns  and to account for the main properties of
 these solutions. The model includes the interaction of the
slowly varying envelope with the small scale of the underlying
pattern solution \cite{ClercFalcon}, well-known as the non-adiabatic
effect \cite{Bensimon,Pomeau}.

\begin{figure}[h!]
\epsfclipon \epsfig{file=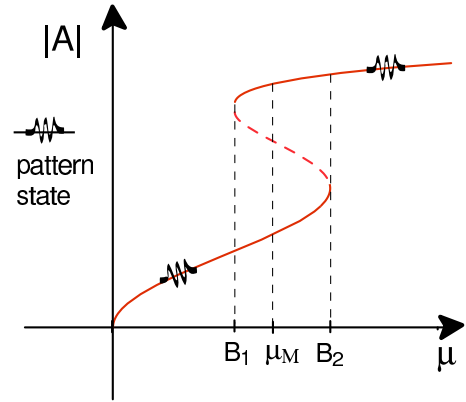,width=6.cm} \caption{ A typical
bifurcation diagram allowing for the appearance of localized peaks:
at a certain value of $\mu$ a secondary subcritical bifurcation
takes place; dashed lines mark the beginning (end) $B_1$ ($B_2$) of
the bistable region and the Maxwell point $\mu_M$.}
\label{Fig-bif-diag}
\end{figure}

Generally, the main ingredient for the appearance of localized peaks
is the coexistence between two spatially periodic states. In order
to give a generic description of such a situation, we consider a
system that exhibits  a sequence of spatial bifurcations as shown in
Fig.\ref{Fig-bif-diag}, that is, the primary bifurcation is
super-critical while the secondary one is of subcritical type. Let
$\vec{u}\left( x;t\right) $ be a vector field that describes the
system under study and satisfies the partial differential equation
\begin{equation}
\partial _{t}\vec{u}=\vec{f}\left( \vec{u},\partial _{x},\left\{ \lambda
_{i}\right\} \right) ,  \label{E-OriginalModel}
\end{equation}
where $\left\{ \lambda _{i}\right\} $ is a set of parameters. For a
critical value of one of the parameters, the system exhibits a
spatial instability at a given wave number $q_c$. Close to this
spatial instability, we use the ansatz
$\vec{u}=A(X,T)e^{iq_cx}\hat{u}+\bar{A}(X,T)e^{-iq_cx}\hat{\bar{u}}+\cdots
$ and the amplitude satisfies \cite{Cross}
\begin{equation}
\partial _{T}A=\mu A-\nu |A|^{2}A+\alpha |A|^{4}A-|A|^{6}A+\partial _{XX}A,
\label{E-Amplitude}
\end{equation}
where $\mu $ is the bifurcation parameter and $\left\{ \nu ,\alpha
\right\}$ control the type of the bifurcation (first or second
order  depending on the sign of these coefficients). Higher-order
terms are ruled out by scaling analysis, since $\nu\sim
\mu^{2/3}$, $\alpha\sim \mu^{1/3}$, $|A|\sim \mu^{1/6}$,
$\partial_t\sim \mu$, $\partial_x\sim \mu^{1/2}$, and  $\mu\ll 1$.
Note that this approach is phase invariant ($A\rightarrow
Ae^{i\varphi }$), but the initial system under study does not
necessarily have this symmetry.

\begin{figure}[h!]
\epsfclipon \epsfig{file=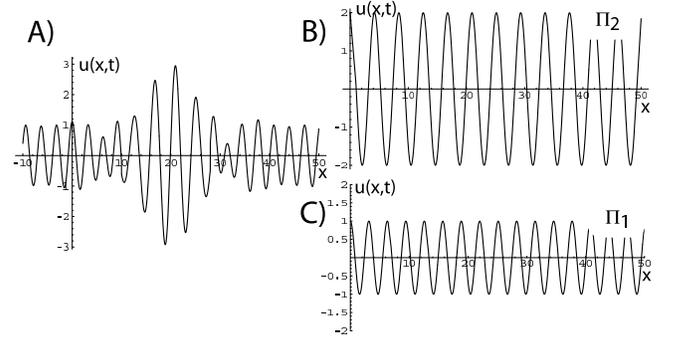,width=8.5 cm} \caption{Localized
state and pattern solutions: a) localized state solution between
pattern state $\Pi_1$ and $\Pi_2$;  b) and c) represent the pattern
solutions $\Pi_1$ and $\Pi_2$ coexisting for the same parameters.}
\label{Fig-ParticleSolution}
\end{figure}

As depicted in Fig.\ref{Fig-bif-diag}, for a given range of
parameter values the system exhibits coexistence between two
different spatially periodic states, each one corresponding to a
homogeneous state for the amplitude equation. The coexistence
region is for $B_{1} < \mu < B_{2}$. The extended
stationary solution of the amplitude equation Eq.(\ref{E-Amplitude}), has
the form ($\partial_tA=0$)
\[
A=R_{o}e^{i \frac{\varepsilon}{R_{o}^{2}}X},
\]
where $\mu -\varepsilon^{2}/R_{o}^{4}-\nu R_{o}^{2}+\alpha
R_{o}^{4}-R_{o}^{6}=0$ and $\varepsilon$ is an arbitrary constant
related to the initial phase invariance. It is worth to note that
in the case of positive $\varepsilon$, the wave number of the
pattern is modified by the inverse of the square amplitude
$R_0^2$, so that patterns with larger amplitude have smaller wave
number. At variance when $\varepsilon$ is negative, the patterns
with large amplitude have smaller wavelength. In
Fig.\ref{Fig-ParticleSolution} are depicted two different patterns
that coexist for the same parameters and the pattern with large
amplitude has smaller wavelength, hence $\varepsilon$ in this case
is negative.

Note that the above amplitude equation is variational and can be
written as
\[
\partial _{t}A=-\frac{\delta {\cal F}
\left[ A , \; \overline A \right] }{\delta \;  \overline{A}},
\]
where
\[
{\cal F}=-\int \left (\mu {|A|^{2} }- \nu{ {|A|^{4}} \over
2}+\alpha { {|A|^{6}} \over 3} -{{|A|^{8}} \over 4}-{\mid \partial
_{X}A \mid^2  } \right ) dx.
\]

For given values of the parameters, the two stable uniform
stationary states of Eq.(\ref{E-Amplitude}) have the same energy,
that is, the system is at the {\it Maxwell point}. Where the front
between the two states does not propagate, that is, the front is
motionless \cite{Collet}. By moving away from the Maxwell point, the
front dynamics is usually characterized by the motion  of the core
of the front, which is defined as the front position with the
largest slope. In order to have a localized solution, we consider
the interaction of two of these motionless fronts close to the
Maxwell point. As a consequence of the asymptotic behavior of the
front at  infinity, the front interaction is attractive, and has the
form \cite{Kawasaki}
\begin{equation}
\dot{\Delta}=-ae^{-\lambda \Delta }+\delta,
\label{E-Interaction}
\end{equation}
where $\Delta $ is the distance between the cores of each front,
$\delta $ is the separation from the Maxwell point ($\mu-\mu_M$),
$\lambda$ characterizes the exponential decay of the front to a
given constant value at infinity, and  $a$ is a positive coefficient
that characterizes the properties of the interaction and is
determined by the form of the front. The Eq.(\ref{E-Interaction})
has an unstable fixed point $\Delta^*=-\ln(\delta/a)/\lambda$, which
is the nucleation barrier between the two homogeneous states. Hence,
the conventional amplitude equation, Eq.(\ref{E-Amplitude}), does
not exhibit stable localized states, due to the scale separation
used to derive the amplitude equation. But near the front's core,
the previous ansatz is no more valid. Indeed, in these locations the
slowly varying envelope $A\left( X,T\right)$ shows oscillations of
the same (or comparable) size as the small scale of the underlying
pattern. This phenomenon is denominated as the non-adiabatic effect
\cite{Bensimon,Pomeau}.

In order to take into account this effect, we modify the amplitude
equation by including the non-resonant terms. Thus,
the amplitude equation becomes
\begin{eqnarray}
\partial _{T}A &=&\mu A-\nu |A|^{2}A+\alpha |A|^{4}A-|A|^{6}A+\partial _{XX}A
\label{E-AmplitudeModify} \nonumber \\
&&+\sum\limits_{m,n \geq
0}g_{mn}A^{m}\bar{A}^{n}e^{-i\frac{q_c(1+n-m)}{\sqrt{\mu }}X}
\end{eqnarray}
where $g_{mn}$ are real numbers of order one. Now the amplitude
equation is parametrically forced in space by the non-resonant
terms.  We note that the ansatz for $\vec{u}$ satisfies the
symmetries $\left\{ x\rightarrow -x,\text{ }A\rightarrow
\bar{A}\right\} $, and $\left\{ x\rightarrow x+x_{o}\text{,
}A\rightarrow Ae^{iq_cx_{o}}\right\}$. Therefore, the envelope
equation also is invariant under this transformation. Instead, the
spatial translation and phase invariance are independent
symmetries of Eq. (\ref{E-Amplitude}).

\begin{figure}[tb]
\epsfclipon \epsfig{file=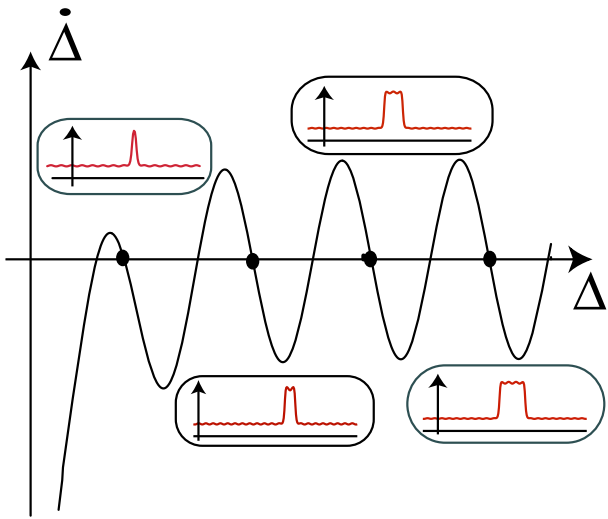,width=8 cm} \caption{Oscillatory
interaction force between two front solutions. The inset figures are
the stable localized patterns observed at the Maxwell points (black
dots), where the interaction changes its sign.}
\label{F-FrontInteraction}
\end{figure}

To understand and illustrate the effect of non-resonant terms we
keep the leading term $n=0$ and $m=2$. Then the amplitude equation
takes the form
\begin{eqnarray}
\partial _{T}A &=&\mu A-\nu |A|^{2}A+\alpha |A|^{4}A-|A|^{6}A+\partial _{XX}A
\label{E-AmplitudeModify} \nonumber \\
&&+\eta A^{2}e^{i\frac{q_c}{\sqrt{\mu }}X.}
\end{eqnarray}
The amplitude is now spatially forced with  frequency $q_c/ 2
\pi\sqrt{\mu }$ and amplitude $\eta \equiv g_{02}$. The spatial
forcing is responsible for the homogenous states becoming a
spatially periodic state.  As a consequence, the front solution
between the spatially periodic states exhibits a pinning range, that
is, the front is motionless for a range of parameter around the
Maxwell point. It is important to remark that the model
(\ref{E-AmplitudeModify}) is the simplest model that exhibits front
solution between two different spatial periodic solutions.

Note that the maxima of the envelope correspond to the maxima of
the initial periodic solution $\vec{u}(t,x)$. In order to obtain
the change of the front interaction as a result of the spatial
forcing, we consider the front solution of the resonant equation

$A_{\pm}(x-x_o)=R_{\pm}(x-x_o)e^{i\int \varepsilon /R_{\pm}^{2}
dx}$, where $R_{\pm}(x-x_o)$ satisfies
\[
\mu R-\nu R^{3}+\alpha
R^{5}-R^{7}+\partial_{xx}R-\frac{\varepsilon^2}{R^{3}}=0,
\]
$x_o$ is the position of the front core and the lower index $+$
($-$) correspond to a front monotonically rising (decreasing). As
the non resonant term is a rapid spatial oscillation, we consider
this term as perturbative-type and use the anstaz
\[
A=A_{+}(x-x_{1}(t))+A_{-}(x-x_{2}(t))-(A_{o,+}-A_{o,-})+\delta W
e^{i\delta\varphi},
\]
in the Eq.(\ref{E-AmplitudeModify}), where
$A_{o,\pm}=R_{o,\pm}e^{i\varepsilon x/R^2_{o,\pm}}$, and $\{
\delta W, \delta\varphi\}$ are small functions, and $R_{o,\pm}$
are the stable equilibrium states of the resonant amplitude
equation (\ref{E-Amplitude}) and $R_{o,+}> R_{o,-}$. We obtain the
following solvability condition for the $\delta W$ function (front
interaction)

\begin{equation}
\dot{\Delta}=-ae^{-\lambda \Delta }+\delta +\gamma \cos \left(
\frac{q_c}{ \sqrt{\mu }}\Delta \right) , \label{E-ModifyInteraction}
\end{equation}
with
\begin{eqnarray}
a &=& \frac{-2\langle 3\mu R_{+}^{2}-5\nu R_{+}^{4}+7 \alpha
R_{+}^{6}-3\varepsilon R_+^{-4}|\partial_{x}R_{+}\rangle}{\langle
\partial_{x}R_{+}|\partial_{x}R_{+}\rangle}, \nonumber \\
\delta &=&\frac{F(R_+)-F(R_-)} {\langle
\partial_{x}R_{+}
 |\partial_{x}R_{+}\rangle}, \nonumber \\
\gamma &=& \frac{\eta \langle \partial_{x}R_{+} | R_{+}^2
\cos\left(\frac{q_c} {\sqrt{\mu}} x \right)  \rangle} {\langle
\partial_{x}R_{+}|\partial_{x}R_{+}\rangle}, \nonumber
\end{eqnarray}
$ F(R)= \mu R^{2}/2- \nu  {R^{4}}/
4+\alpha  R^{6} / 6 -R^{8}/8+ 2\varepsilon^2 / R^{2}$, and
$\langle f |g\rangle\equiv \int_{-\infty}^{\infty}f(x)g(x)dx $.

As a consequence of the spatial forcing the interaction of two
fronts close to the pinning range, Eq.(\ref{E-ModifyInteraction}),
has an extra term and now alternates between attractive and
repulsive forces.  It is important to remark that $\gamma $ is a
parameter exponentially small, proportional to $\eta$, and is of
order $\delta $, i.e. the source of the {\it periodical force} is
the spatial forcing in the Eq.( \ref{E-AmplitudeModify}).
Therefore, close to the Maxwell point the system exhibits a family
of equilibrium points, $d\Delta / dt=0$. Each equilibrium point
correspond to a localized solution nucleating over a pattern
state, we call these solutions {\it localized patterns}. The
lengths of localized patterns are multiple of a basic length,
corresponding to the shortest localized state. We term these
shortest states as {\it localized-peaks}, as these solutions
correspond to the experimental observations reported in
\cite{Bortolozzo2005}.
In Fig.\ref{F-FrontInteraction}, it is depicted the
front interaction and the family of equilibrium points.

Due to the oscillatory nature of the front interaction, which
alternates between attractive and repulsive forces (cf.
Fig.\ref{F-FrontInteraction}), we can deduce the dynamical
evolution and bifurcation diagram of localized patterns. By
decreasing $\delta$ or increasing $\eta$, the family of localized
patterns disappears by successive saddle-node bifurcations and
only localized peaks survive. The mechanism for localized peak
appearance is related to the fact that the spatial forcing is
nonlinear. Indeed, it is proportional to the square of the pattern
amplitude.

Since the amplitude of spatial forcing for the upper branch is
larger than that for the lower branch, then the patterns with large
modulus have large spatial amplitude oscillations around the
equilibrium state of unperturbed amplitude equation (cf.
Fig.(\ref{Fig-ParticleSolution})). Thus, for a given critical, and
small, value of the forcing the pattern with high  magnitude becomes
unstable, because this state collides with the unstable pattern
state. In Fig. (\ref{Fig-bif-diag}), this unstable state is
represented by the dashed line. Hence, the minima of the pattern
with high  magnitude reach the pattern with lower magnitude, and
 a saddle-node bifurcation of the spatial periodic solution
gives rise to the appearance of a localized peak. Because of this
mechanism, localized patterns with a size larger than the shortest
length are not robust phenomena. In fact, the typical behavior
observed in the experiments is the appearance of localized-peaks
\cite{Bortolozzo2005}.

\begin{figure}[h!]
\epsfclipon \epsfig{file=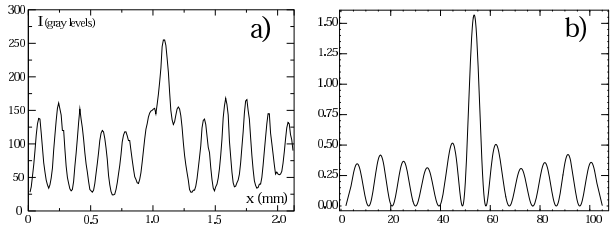,width=8 cm} \caption{ a) Intensity
profile of a one-dimensional localized peak in the LCLV experiment;
b)  $\| \vec{u}\|^2$ numerical profile in the presence of a
localized peak.} \label{profile}
\end{figure}

In Fig.\ref{profile}a, it is shown a localized peak profile 
recorded in the Liquid-Crystal-Light-Valve (LCLV) experiment
\cite{Bortolozzo2005}. In order to directly compare with the model,
we have performed one-dimensional experiments by inserting a
rectangular slit in the optical feedback loop. The slit transverse
size is approximately $100$ $\mu m$ whereas, for the parameters set
in the experiment,  the size of localized peaks is around $350$ $\mu
m$.   A similar profile can be numerically obtained for $\|
\vec{u}\|^2=\mid A\mid^2 \cos(q_c x)$, as shown in
Fig.\ref{profile}b.

In conclusion, we have presented an unifying description  of
localized peaks, which are large amplitude peaks nucleating over a
lower amplitude pattern. We have derived a spatially forced
amplitude  equation and shown that localized peaks are a generic
class of behavior appearing whenever a pattern forming system
exhibits coexistence between two spatially periodic states. The
front solution that connects the two different pattern states
exhibit a locking phenomena, that is, it is motionless for a range
of parameter. We have obtained the front interaction and from this
interaction we have deduced the family of localized solutions. We
have shown that, as a consequence of the nonlinear nature of the
forcing, localized patterns with a size larger than the shortest
length are not robust phenomena, so that only localized peaks are
stable at long times and for a wide range of parameters. We have
shown a good qualitative agreement with the experimental
observations for a LCLV system and we expect similar phenomena to be
observed in other pattern forming systems, provided they present
bistability between two different spatial structures. Note that
pinning of localized structures on periodic arrays has recently been
reported for a fixed grid \cite{Alexander}. Localized peaks can be
seen as a generalization of this case, when the pinning lattice is
spontaneously generated by the system itself.

The simulation software {\it DimX}, developed at INLN, has been
used for  all the numerical simulations presented in this paper.
The authors thanks the support of ECOS-CONICYT collaboration
program. M.G. C. acknowledges the financial support of FONDECYT
project 1051117, and FONDAP grant 11980002. R.R. acknowledges
financial support from Beca Presidente de la Rep\'ublica of the
Chilean Government.


\begin{references}

\bibitem{Eschenfelder}  H.A. Eschenfelder, {\it Magnetic Bubble Technology }
(Springer Verlag, Berlin 1981).

\bibitem{Oswald}  S. Pirkl, P. Ribi\`ere and P. Oswald, Liq. Cryst. {\bf 13},
413 (1993).

\bibitem{Astrov97}  Y.A Astrov and Y.A. Logvin, Phys. Rev. Lett. {\bf 79},
2983 (1997).

\bibitem{Swinney}  K-Jin Lee, W. D. McCormick, J.E. Pearson and H.L.
Swinney, Nature {\bf 369}, 215 (1994).

\bibitem{Edwards}  W.S. Edwards and S. Fauve, J. Fluid Mech. {\bf 278}, 123
(1994).

\bibitem{Melo}  P.B. Umbanhowar, F. Melo and H.L. Swinney, Nature {\bf 382},
793 (1996).

\bibitem{Heinrichs}  R. Heinrichs, G. Ahlers and D.S. Cannell, Phys. Rev. A
{\bf 35}, R2761 (1987); P. Kolodner, D. Bensimon and C.M. Surko, Phys. Rev.
Lett. {\bf 60}, 1723 (1988).

\bibitem{Newell}  D.W. Mc Laughlin, J.V. Moloney and A.C. Newell, Phys. Rev.
Lett. {\bf 51}, 75 (1983).

\bibitem{Cross}  M. Cross and P. Hohenberg, Rev. Mod. Phys. {\bf 65}, 851
(1993).

\bibitem{Coullet2002}  P. Coullet, C. Riera and C. Tresser,
Phys. Rev. Lett. {\bf 84}, 3069 (2000).

\bibitem{VanSaarlos} W. van Saarloos and P.C. Hohenberg, Phys. Rev. Lett.
{\bf 64}, 749 (1990).

\bibitem{Bortolozzo2005} U. Bortolozzo, R. Rojas  and S. Residori,
Phys. Rev. E {\bf 72}, 045201(R) (2005).

\bibitem{Fineberg}  H. Arbell and J. Fineberg,
Phys. Rev. Lett. {\bf 85}, 756 (2000).

\bibitem{atomic} Yu. A. Logvin, B. Sch\"apers and T. Ackemann,
Phys. Rev. E {\bf 61}, 4622 (2000).

\bibitem{Lionel} L. Gil, private communication.

\bibitem{ClercFalcon}  M.G. Clerc and C. Falcon, Physica A \textbf{356}, 48 (2005).

\bibitem{Bensimon}  D. Bensimon, B.I. Shraiman, and V.Croquette, Phys. Rev.
A {\bf 38}, R5461 (1988).

\bibitem{Pomeau}  Y. Pomeau, Physica D {\bf 23}, 3 (1986).

\bibitem{Collet} P. Collet, J.P. Eckmann, {\it Instabilities and Fronts in
Extended systems}, (Princeton University Press, New Jersey, 1990).

\bibitem{Kawasaki}  K. Kawasaki, and T. Ohta, Physica A {\bf 116}, 573
(1982).

\bibitem{Alexander} T.J. Alexander, A. A. Sukhorukov, and Y. S. Kivshar,
Phys. Rev. Lett. {\bf 93}, 063901 (2004).





\end{references}
\end{document}